\documentclass[aps,pra,superscriptaddress,amsmath,amssymb,preprintnumbers,twocolumn]{revtex4}
\usepackage{amssymb}
\usepackage{graphicx}
\usepackage{bm}
\usepackage{color}
\usepackage{cancel}
\usepackage{slashed}
\usepackage{subfigure}

\newcommand{\Ignore}[1]{}

\newcommand{\Ket}[1]{\left\vert #1\right\rangle}
\newcommand{\Bra}[1]{\left\langle #1\right\vert}

\newcommand{\BraKet}[2]{\left\langle#1\vert #2\right\rangle}
\newcommand{\KetBra}[2]{\left\vert#1\right\rangle\left\langle#2\right\vert}

\newcommand{\ii}{{\mathrm{i}}}
\newcommand{\ee}{\mathrm{e}}

\begin{document}

\title{Steepest Entropy Ascent for Two-State Systems with Slowly Varying Hamiltonians}

\author{Benedetto Militello}
\address{Dipartimento di Fisica e Chimica, Universit\`a degli Studi di Palermo, Via Archirafi 36, I-90123 Palermo, Italy}
\address{I.N.F.N. Sezione di Catania, Italy}

\begin{abstract}
The Steepest Entropy Ascent approach is considered and applied to few-state systems. When the Hamiltonian of the system is time dependent, the principle of maximum entropy production can still be exploited; arguments to support this fact are given.  
In the limit of slowly varying Hamiltonians which allows for the adiabatic approximation for the unitary part of the dynamics, the system exhibits significant robustness to the thermalization process. Specific examples such as a spin in a rotating field and a generic two-state system undergoing an avoided crossing are considered.
\end{abstract}

\maketitle

\section{Introduction}\label{sec:introduction}

It is a well established knowledge that every physical system naturally approaches an equilibrium condition, which is a microcanonical, canonical or grand canonical state, depending on the constraints the system is subjected to. In spite of the great success of statistical mechanics in the study of systems at equilibrium, to the point of reproducing all the behaviors described by a phenomenological theory such as the thermodynamics~\cite{ref:TolmanBook, ref:LandauBook, ref:HuangBook}, an important and very old issue is to properly describe the approach to equilibrium and deeply understand the mechanisms that are responsible for it. Stated in a different way, it is important to describe systems out of equilibrium. A widely used approach to this problem is that of open quantum systems~\cite{ref:PetruBook,ref:GardinerBook}. Sometimes specific hypotheses about the system-environment interaction are considered~\cite{ref:Tasaki}, while in other cases less common techniques of derivation of the relevant master equations are considered, as for example the approach based on Hilbert space averages~\cite{ref:Gemmer2003,ref:Gemmer2005,ref:Gemmer2006}. Exploitation of Quasi Normal Modes is another way to take into account the environment~\cite{ref:QNM}. Evolutions of open quantum systems exhibit intriguing features such as non-Markovianity, which has been extensively studied~\cite{ref:nonMarkov1,ref:nonMarkov2,ref:nonMarkov3,ref:nonMarkov4}, or counterintuitive behaviors such as the dissipation-induced or temperature-induced quantum Zeno effect and related phenomena~\cite{ref:Maniscalco,ref:Kurizki,ref:Militello}.

Though successful in the description of systems interacting with an environment, the theory of open quantum systems cannot describe situations where the system thermalizes in spite of being seemingly non-interacting with its surroundings. Over the decades, the idea of an intrinsic tendency of a quantum system toward the equilibrium condition governed by some principle has been introduced, from the seminal works of Onsager~\cite{ref:Onsager} to the Prigogine's principle of minimum entropy production~\cite{ref:Prigogine}, to the the Maximum Entropy Production Principle (MEPP)~\cite{ref:Martyushev}. A method strongly related to the MEPP is based on the Steepest Entropy Ascent (SEA) approach~\cite{Beretta1984,ref:Beretta2006,ref:Beretta2007,ref:Beretta2009,ref:Beretta2014}, on which we will focus in this paper in the spirit of exploring intriguing properties coming from the relevant mathematical apparatus. 

SEA approach consists in the assumption that at every instant of time the system is \lq pushed\rq\, in the direction of steepest entropy ascent compatible with the constraints imposed on the system, so that for a time-independent Hamiltonian it evolves toward the unique state which maximizes the entropy for the initial values of the constraints. With a suitable {\it ansatz} about the structure of the equation for the system density operator, the generator of the evolution can be systematically obtained. Microcanonical and canonical states are easily proven to be stationary states of the relevant master equations. Though the tendency to a Gibbs state is a shared property --- it is actually the basic requirement of any theory that is claimed to describe the approach to equilibrium ---, the dynamical behaviors obtained from different techniques can significantly differ~\cite{ref:Smith2016}.

In a spirit of exploration of the SEA approach~\cite{ref:footnote}, it is worth wondering whether it can be applied to time dependent Hamiltonians. In fact, there is a great interest toward systems governed by time dependent Hamiltonians, since they can be used to suitably manipulate the quantum state of the system. In particular, adiabatic or quasi adiabatic evolutions are important in the study of geometric phases, and of such processes as Landau-Zener transitions~\cite{ref:LZ1,ref:LZ2} and Stimulated Raman Adiabatic Passage~\cite{ref:STIRAP}. 

In this paper we will explore the problem of deriving the relevant SEA equation governing the system when it is subjected to a time dependent Hamiltonian, focusing on Two-State Systems (TSSs) with slowly-varying Hamiltonians.
In Section~\ref{sec:SEA_General} general aspects of the SEA approach are recalled, from the technique to derive the master equation to the identification of stationary states. In Sec.~\ref{sec:SEA_FSS} the SEA approach is applied to TSSs, pointing out very specific peculiarities implied by the constraints that have to be imposed to the SEA part of the relevant master equation. In Sec.~\ref{sec:SEA_TD} we deal with the problem of SEA approach for time-dependent Hamiltonians: we discuss the general approach and then investigate general properties that can be attributed to adiabatic evolutions. Specific examples are discussed. Finally, in Sec.~\ref{sec:conclusion} we give some conclusive remarks.

\section{Steepest Entropy Ascent Approach}\label{sec:SEA_General}

Steepest Entropy Ascent (SEA) approach is an axiomatic way to study the out-of-equilibrium dynamics of quantum systems, which inevitably approach an equilibrium state. Since a system naturally evolves toward the state with maximum entropy compatible with the relevant constraints (probability preservation, conservation of energy, etc.), it is reasonable to assume that {\it at every instant of time the system is pushed toward such a configuration that maximizes the entropy, provided the conservation of the relevant constants of motion}.

The SEA principle~\cite{ref:Beretta2009,ref:Beretta2014} states that the system tends toward its equilibrium state following a path always tangent to the direction of the entropy gradient with respect to a phenomenologically attributed metric field defined over the entire nonequilibrium state space. In addition, the strength of such tendency is also attributed phenomenologically to the system through the specification of a relaxation time functional defined over the entire non-equilibrium state space.

This approach is aimed at providing an alternative to standard statistical physics or to the theory of open quantum systems as a bridge between quantum mechanics and thermodynamics, proposing a suitable extension of quantum mechanics (for a discussion see for example Sec.1.2 of Ref.~\cite{ref:Beretta2009}). In fact, it introduces a microscopic dynamical law which reduces to the Liouville-von Neumann equation for perfectly pure states (hence reproducing the standard quantum mechanics) and which introduces tendency to the thermodynamic equilibrium for non-pure states.

In order to derive the proper master equation that governs the time evolution of a quantum system, we need to assume a general structure for the equation. The simplest form that one can assume is the following one, which guarantees the Hermiticity at every instant of time:
\begin{equation}\label{eq:Hermiticity}
\dot\rho = \rho E + E^\dag \rho  \,, 
\end{equation} 
where the operator $E$ can be a nonlinear one, meaning that it can depend on $\rho$  itself, being $E(\rho)$.

The operator $E$ can be determined by imposing that it induces both the unitary evolution and a dissipative dynamics which preserves population, energy expectation value, and the expectation values of other possible constants of motion. Then we get $E(\rho)=\ii H + E_D(\rho)$, where $E_D$ induces the non-unitary part of the evolution. 
As summarized in appendix \ref{app:SEAME_Canonical}, assuming the simple structure of Eq.~\eqref{eq:Hermiticity}, explicit expression of the time-derivative of a generic observable, as well as of the entropy functional, naturally lead to the introduction of a suitable real scalar product in the operator space (which in turn can be connected to the Fisher-Rao metric for density operators, as discussed in Refs.~\cite{ref:Beretta2009,ref:Beretta2014}). With the help of this tool, the expression of $E_D$ is found by considering the gradient of the entropy functional and removing those \lq components\rq\, that can produce variations of quantities that must be conserved.

When probability and energy conservations are the only constraints, the form of the master equation is the following:
\begin{subequations}
\begin{eqnarray}\label{eq:SEAME_Canonical}
\nonumber
\dot\rho &=& -\ii[H, \rho] - \gamma(\rho) \left( \rho\log\rho - \mu(\rho) \rho + \nu(\rho) \{ \rho, H\} \right)\,,\\
\end{eqnarray}
with
\begin{eqnarray}
\label{eq:SEAME_Canonical_Def1}
\mu(\rho) &=& \frac{1}{\sigma^2_{H}}(s(\rho) \langle H^2\rangle_\rho - \langle H\rangle_\rho \langle\log\rho H\rangle_\rho) \,, \\
\label{eq:SEAME_Canonical_Def2}
\nu(\rho) &=& \frac{1}{2\sigma^2_{H}}(s(\rho)\langle H\rangle_\rho - \langle\log\rho H\rangle_\rho) \,,
\end{eqnarray} 
where
\begin{eqnarray} 
\langle X \rangle_\rho &=& \mathrm{tr}(\rho X) \,, \\
\sigma^2_{H} &=& \langle H^2\rangle_\rho - \langle H\rangle_\rho^2 \,,
\end{eqnarray} 
\end{subequations}
and where $[A, B]$ and $\{A, B\}$ denote the standard commutator and anti-commutator operations. The functional $\gamma(\rho)$ determines the strength of the tendency toward the equilibrium, and is determined phenomenologically. In the following we will always assume this rate to be constant: $\gamma(\rho)=\gamma$.

It is worth mentioning that this result corresponds to the following:
\begin{eqnarray} 
E_D(\rho) = - \gamma(\rho) \left(\frac{1}{2}\log\rho - \frac{1}{2} \mu(\rho) {I} + \nu(\rho) H \right)\,,
\end{eqnarray} 
with ${I}$ the identity operator on the relevant Hilbert space.

It is well known~\cite{ref:Beretta2007,ref:Beretta2009} and straightforward to prove (see appendix~\ref{app:CanonicalState}) that this equation admits every canonical state $\omega(\beta) = e^{-\beta H} / \mathrm{tr}(e^{-\beta H})$ as a stationary state, for every $\gamma(\rho)$. Moreover, every restriction of any of such density operators to a subspace generated as direct sum of eigenspaces of the Hamiltonian is a stationary state as well: $\omega_{\hat{P}}(\beta) = \hat{P} e^{-\beta H} / \mathrm{tr}(\hat{P} e^{-\beta H})$, with $\hat{P}$ a projection operator such that $[H, \hat{P}]=0$, is a stationary state for the SEA master equation in \eqref{eq:SEAME_Canonical}. This naturally follows from the fact that when $\rho=\hat{P}\rho$ both $\rho\log\rho$ and, obviously, $\rho$ itself do not have matrix elements out of the subspace corresponding to $\hat{P}$, and, in addition, because of the commutation with the Hamiltonian, neither the commutator nor the anti-commutator turns out to be out of the subspace identified by $\hat{P}$. As a particular case, every eigenstate of the Hamiltonian is a stationary state for the SEA master equation. By the way, every exactly pure state is insensitive to $E_D$ and evolves unitarily according to $\dot\rho = -\ii [H, \rho]$. In fact, for a pure state the operator $\rho\log\rho$ is the null operator and, as a consequence, both the functionals $s(\rho)$ and $\langle\log\rho H\rangle_\rho$ are zero, which in turn implies $\nu(\rho)=0$ and $\mu(\rho)=0$. Nevertheless, even a very small deviation from being pure makes the state (whether close to an Hamiltonian eigenstate or not) trigger the SEA evolution and inevitably brings the system toward the canonical state. 

A natural question rises: can the system evolve toward any canonical state or is it forced to choose a specific value of $\beta$? The answer is simpler than it can seem, since, by construction, the master equation conserves the energy and therefore the system is pushed toward the canonical state whose temperature is such that the average energy turns out to be equal to the average energy of the initial state: 
\begin{equation}\label{eq:EnergyConstraint}
\mathrm{tr}(\omega(\beta)H) = \mathrm{tr}(\rho(0)H)\,.
\end{equation}

Summarizing, we have a master equation which conserves Hermiticity (because of the structure in \eqref{eq:Hermiticity}), normalization and energy (by construction, imposing the relevant constraints). Such master equation describes a non-unitary pushing toward the state with highest entropy compatible with the constraints; such pushing competes with the unitary part of the evolution traceable back to the commutator with the Hamiltonian in the master equation. The equilibrium state resulting from such competition is a Gibbs state with a temperature connected to the initial energy of the system. It is important to observe here that, from the SEA point of view, the density operator whose dynamics is governed by the master equation \eqref{eq:SEAME_Canonical} is not obtained by tracing over the degrees of freedom of the environment as in the theory of open quantum systems: since the system is assumed to be closed, only its degrees of freedom are considered and the non-unitary part of the dynamics is postulated from the beginning.

\section{SEA evolutions of Two-State Systems}\label{sec:SEA_FSS}

Generally speaking, SEA master equations of the form in \eqref{eq:SEAME_Canonical} tend to kill the coherences between eigenstates of the Hamiltonian and to modify populations in such a way to make them approach the canonical distribution. Nevertheless, when the system is a two-state one it exhibits a peculiarity, which is the conservation of the populations of the eigenstates of the Hamiltonian (namely, the diagonal matrix elements of the density operator in the basis of the eigenstates of the Hamiltonian). 

Consider a TSS whose Hamiltonian eigenvalues are addressed as $\epsilon_0$ and $\epsilon_1$, corresponding to the eigenstates $\Ket{0}$ and $\Ket{1}$. After introducing the populations $p_j=\mathrm{tr}(\rho\Ket{j}\Bra{j})$ satisfying $p_0+p_1=1$, the average energy of the system can be expressed as $U=p_0\epsilon_0 + (1-p_0)\epsilon_1$, which allows for a single realization of the energy value $U$ in terms of the populations of the eigenstates, and, given an energy value, the population of the state $\Ket{0}$ is unambiguously determined as $p_0=(\epsilon_1-U)/(\epsilon_1-\epsilon_0)$, as well as $p_1=1-p_0$. This simple fact, concomitant with the constraint of energy conservation, implies that the two populations cannot change at all during the SEA evolution (of course, coherence can change).
This property is peculiar of TSSs, because in the presence of more than two states a given value of energy can be obtained usually in several (infinite) ways, with the exception of those cases where the average coincide with the minimum or maximum energy value. This means that out of the case of a single TSS, populations of the energy eigenstates will be rearranged, provided the probability normalization, the constraint in \eqref{eq:EnergyConstraint} and other possible constraints, are satisfied. Two-state systems with time-independent Hamiltonians instead allow only for changes of the coherence between the two eigenstates of the Hamiltonian.

The dynamics of the TSS is assumed to be generated by the master equation \eqref{eq:SEAME_Canonical} associated to the following Hamiltonian, here written in the basis of its eigenstates $\{\Ket{1},\Ket{0}\}$ ($\hbar=1$):
\begin{eqnarray} 
H_2 = \left(
\begin{array}{cc}
\epsilon & 0 \\ 
0 & 0 
\end{array}
\right)\,.
\end{eqnarray} 

\begin{figure}
\subfigure[]{\includegraphics[width=0.45\textwidth, angle=0]{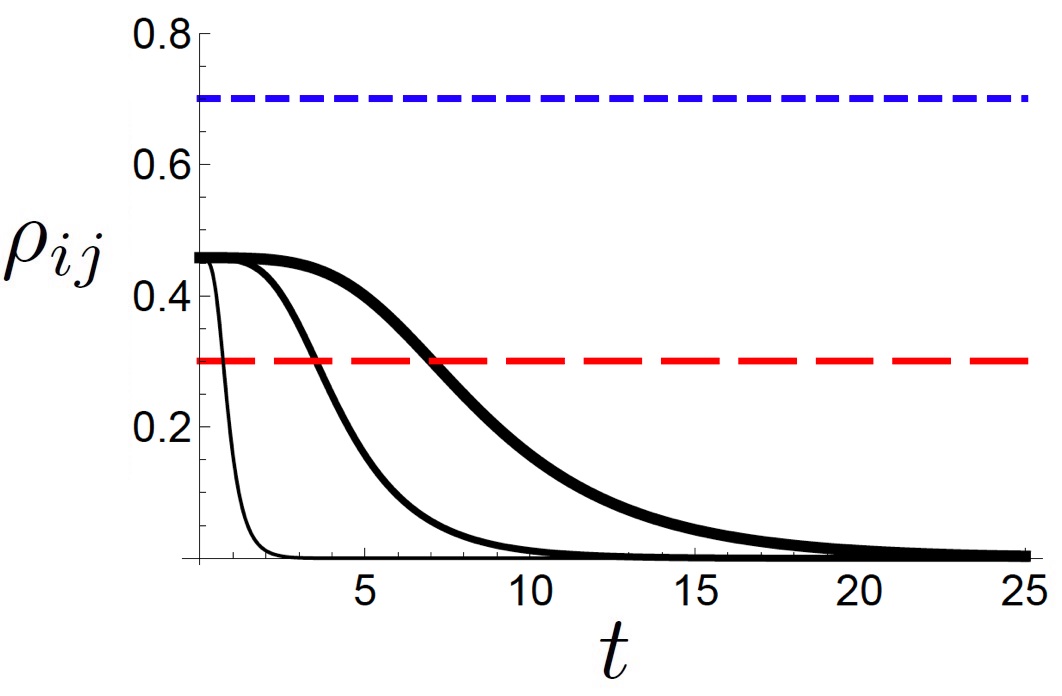}}
\subfigure[]{\includegraphics[width=0.45\textwidth, angle=0]{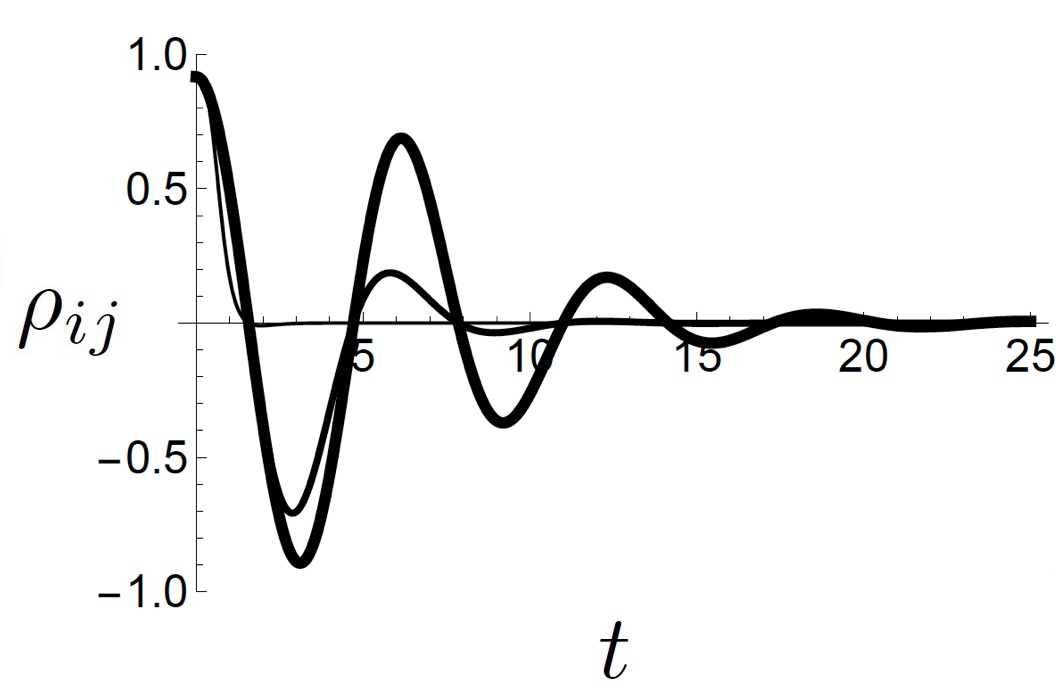}}
\subfigure[]{\includegraphics[width=0.45\textwidth, angle=0]{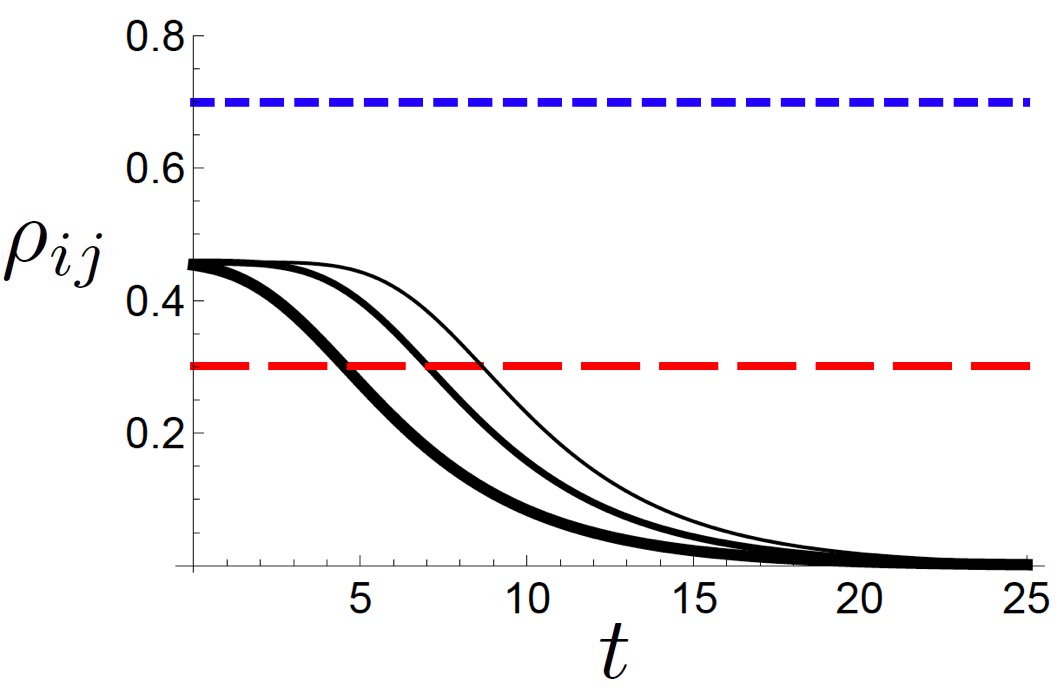}}
\caption{Matrix elements of the density operator of a TSS as a function of time (in units of $1/\epsilon$), when the initial state is $\rho(0)=(1-\lambda)\Ket{\psi_0}\Bra{\psi_0}+(\lambda/2) I$ with $\Ket{\psi_0}=\sqrt{0.7}\Ket{1}+\sqrt{0.3}\Ket{0}$. The matrix elements are meant with respect to the basis of the Hamiltonian eigenstates. In (a) the plotted quantities are: $\rho_{11}$ (blue dashed line), $\rho_{00}$ (red long dashed line), $|\rho_{01}|$ for $\gamma/\epsilon=0.25$ (bold black solid line), $|\rho_{01}|$ for $\gamma/\epsilon=0.5$ (black solid line) and $|\rho_{01}|$ for $\gamma/\epsilon=2.5$ (thin black solid line). The curves describing the two populations $\rho_{00}$ and $\rho_{11}$ (straight lines) for the three values of $\gamma/\epsilon$ considered perfectly coincide; for the three values of $\gamma/\epsilon$ it has been used $\lambda=10^{-4}$. 
Plots of the quantity $\langle \sigma_x \rangle$ are reported in (b) for the three values of $\gamma/\epsilon$ previously considered: $\gamma/\epsilon=0.25$ (bold black solid line), $\gamma/\epsilon=0.5$ (black solid line) and $\gamma/\epsilon=2.5$ (thin black solid line). In (c) the plotted quantities are: $\rho_{11}$ (blue dashed line), $\rho_{00}$ (red long dashed line), $|\rho_{01}|$ for $\lambda=10^{-2}$ (bold solid black line), $|\rho_{01}|$ for $\lambda=10^{-4}$ (black solid line), $|\rho_{01}|$ for $\lambda=10^{-6}$ (thin solid black line); for all values of $\lambda$, $\gamma/\epsilon=0.25$ has been considered.
} \label{fig:2SS}
\end{figure}

In the previous section it has been stated that, in spite of the nonlinear form of the master equation, when the system dynamics starts with a perfectly pure state the evolution is perfectly unitary, because the operator $\rho\log\rho$ is the null operator and all the SEA contributions vanish. Therefore, in order to make the SEA evolution start, we will always assume that the system is initially in a pure state very slightly perturbed: $\rho(0)=(1-\lambda)\Ket{\psi_0}\Bra{\psi_0}+ (\lambda/2) I$, with $\lambda \ll 1$. (The values of $\lambda$ are declared in the relevant captions.)

In Fig.~\ref{fig:2SS}a an example of SEA evolution for a TSS prepared in an almost-pure stare ($\lambda=10^{-4}$) is plotted. The populations $\rho_{00}$ and $\rho_{11}$ do not change, while the coherence is completely destroyed as the equilibrium state is approached. Three different values of the parameter $\gamma$ are considered, showing, as expected, that higher values of $\gamma$ imply faster decay of the off-diagonal matrix element $\rho_{01}$. In the three cases analyzed, an initial {\it plateau} followed by a relaxation process is evident for the coherences. Fig.~\ref{fig:2SS}b shows the corresponding behaviors of $\langle \sigma_x \rangle \equiv \mathrm{tr}(\rho\sigma_x)$, with $\sigma_x=\Ket{1}\Bra{0}+\Ket{0}\Bra{1}$, which are damped oscillations. Higher values of $\gamma$ correspond to more rapidly decaying oscillations; the curve related to the highest value of $\gamma$ shows a zero amplitude reached before the first oscillation can occur. In Fig.~\ref{fig:2SS}c three different values of $\lambda$ ($10^{-2}$, $10^{-4}$, $10^{-6}$) are considered, and it is well visible that the higher the purity of the initial state is, the slower the approach to equilibrium is. In all these cases (those in \ref{fig:2SS}a and \ref{fig:2SS}c), the behavior of the coherence for the non-degenerate TSS here considered resembles that of the populations of a degenerate TSS discussed in detail in Ref.~\cite{ref:Beretta2006}.

It is worth observing that the temperatures characterizing the equilibrium states of Fig.~\ref{fig:2SS} are all negative (and very close to each other, with very small differences due only to different values of $\lambda$ and hence of $p_0$). Negative temperatures are not surprising in this context, since we are dealing with an upper bounded Hamiltonian~\cite{ref:Ramsey1956,ref:GheorghiuSvirschevski2001}, which is the case of every Hamiltonian acting on a finite-state Hilbert space. Generally speaking, for a TSS negative and positive temperatures can characterize the equilibrium state, depending on whether the condition $p_0>0.5$ (which implies $p_0>p_1=1-p_0$) is satisfied (positive temperature) or not (negative temperature).

\section{Time-Dependent Hamiltonian Models}\label{sec:SEA_TD}

A question about Steepest Entropy Ascent approach that could raise at this point is whether it is exploitable also when the Hamiltonian is time dependent.  In several papers SEA approach in the presence of time-dependent  Hamiltonians has been considered \cite{ref:Beretta1985a,ref:Beretta1985b,ref:Beretta2007}. Actually, no difference has been introduced by the Author of such papers between time-dependent and time-independent cases. Nevertheless, we think that some comments are due when non stationary Hamiltonians are considered, which we will provide in the next subsection. In the subsection \ref{sec:sub_adiabaticity} we instead consider a special class of time-dependent Hamiltonians, that is the slowly varying ones allowing for the adiabatic approximation. In fact, we will show that adiabatic evolutions are pretty insensitive to the action of the non-unitary part of the dynamics related to $E_D(\rho)$.

\subsection{General Framework}\label{sec:SEA_TDGeneral}\label{sec:sub_general_frame}

The master equation in \eqref{eq:SEAME_Canonical} is derived from the Steepest Entropy Ascent approach with specific constraints, which are probability and energy conservation. When the Hamiltonian is time dependent, the energy (i.e., the expectation value of the Hamiltonian operator) is not conserved even in the absence of any form of dissipation or incoherent dynamics. Therefore, one could wonder whether the component of the entropy gradient that can change the Hamiltonian has to be removed or not, in this case. In spite of the seemingly need to relax this constraint, one can observe that if we consider a time scale $\tau$ much smaller than the time scale of the Hamiltonian variations, in a time window $(t, t+\tau)$ the system does not distinguish a time dependent Hamiltonian form a time independent one. This naturally leads at assuming that at each instant of time the conservation of the average value of the energy has to required.

By applying the principle of maximum production of entropy and the constraints of probability preservation and conservation of the Hamiltonian time by time, we get the time-dependent counterpart of \eqref{eq:SEAME_Canonical}:
\begin{eqnarray}\label{eq:SEAMETDl}
\nonumber
\dot\rho &=& -\ii[H(t), \rho] \\
&-& \gamma(\rho) \left( \rho\log\rho - \mu(\rho) \rho + \nu(\rho) \{ \rho, H(t)\} \right)\,,
\end{eqnarray}
where the parameters $\gamma(\rho)$,  $\mu(\rho)$,  $\nu(\rho)$ are defined as in \eqref{eq:SEAME_Canonical_Def1} and \eqref{eq:SEAME_Canonical_Def2}.
This structure of master equation is exactly the one considered in Refs. \cite{ref:Beretta1985a,ref:Beretta1985b,ref:Beretta2007}. 

It is the case to observe that the generator of the non unitary dynamics in this master equation is invariant under Hamiltonian rescaling. Indeed, $H \rightarrow \sigma H$ implies $\mu(\rho) \rightarrow \mu(\rho)$,  $\nu(\rho) \rightarrow \sigma^{-1}\nu(\rho)$,  $\{\rho, H\} \rightarrow \sigma \{\rho, H\}$, which in turn imply invariance of the terms $\rho E_D$ and $E_D^\dag\rho$. Therefore, in the presence of a time dependent Hamiltonian of the form $H(t)=\sigma(t) H_0$ the part of the master equation describing the tendency toward the equilibrium state remains invariant.

\subsection{Adiabaticity}\label{sec:SEA_Adiabatic}\label{sec:sub_adiabaticity}

The master equation in \eqref{eq:SEAMETDl} is general and applies to all kinds of time-dependent Hamiltonians and in particular to the slowly varying ones, on which we will focus in this subsection.
In such case, the simple Hamiltonian evolution turns out to be essentially a mapping of the Hamiltonian eigenstates at the initial time, say $t=0$, to the eigenstates of the Hamiltonian at the generic time $t$: $U(t)\Ket{\phi_n(0)} \approx e^{\ii\alpha_n(t)} \Ket{\phi_n(t)}$, with $H(t)\Ket{\phi_n(t)}=\epsilon_n(t)\Ket{\phi_n(t)}$, $\alpha_n(t)=\int_0^t (-\ii \epsilon_n(s)-\langle\phi_n(s)|\dot\phi_n(s)\rangle)\mathrm{d}s$ and $U(t)$ the evolution operator generated by $H(t)$. In order to guarantee adiabaticity when a generic Hamiltonian is considered, the very well know condition  $|\Bra{\phi_n}\dot{H}\Ket{\phi_m}|/(\epsilon_n-\epsilon_m)^2 \ll 1$ has to be fulfilled for every $m\not=n$~\cite{ref:Messiah, ref:Militello2011, ref:Wilczeck}. (See Appendix \ref{app:Adiabaticity} for details.) 

The adiabatic mapping is intrinsically approximated~\cite{ref:Messiah, ref:Militello2011, ref:Wilczeck}, and the perfect adiabatic following of the eigenstates of the Hamiltonian governing the system can never be reached. Nevertheless, the slower the Hamiltonian change the closer to a perfect adiabatic evolution the unitary time evolution is. Of course, perfect adiabatic following of the eigenstates of an Hamiltonian $H(t)$ can be obtained when the system is governed by an Hamiltonian $H'(t)$ suitably related to $H(t)$, which is the essence of shortcuts to adiabaticity~\cite{ref:Shortcuts1,ref:Shortcuts2,ref:Shortcuts3}.

We have previously observed that the eigenstates of the Hamiltonian (in the time-independent Hamiltonian case) are stationary states of the relevant SEA master equation. Moreover, if the Hamiltonian vary slowly enough the generator of the unitary evolution is responsible, with a good approximation, for an adiabatic following of each eigenstate of the Hamiltonian. Therefore, if the system starts in an eigenstate of the Hamiltonian at the initial time and if such an Hamiltonian is slowly varying, then it is reasonable to expect that the system will remain in the relevant instantaneous eigenstate of the Hamiltonian, because neither the $-\ii[H(t), \rho]$ term will produce an abandon of the eigenstate nor the $\rho E_D + E_D^\dag\rho$ term will realize a significant pushing toward the canonical state associated to $H(t)$. 
By taking into account these facts, one can expect robustness of the adiabatic evolution of an eigenstate of the Hamiltonian against the SEA contributions of the master equation.

{\it Rotating Field} --- In order to better demonstrate the validity of this analysis we will consider a very simple model consisting of a TSS subjected to a periodic time-dependent Hamiltonian. It can represent a spin in a rotating magnetic field or a two-state atom interacting with a classical electric field off-resonant with the atomic frequency, and its matrix form, in the basis of the bare states $\{\Ket{1}, \Ket{0}\}$, is: 
\begin{eqnarray}
H_\mathrm{rot}(t) = \Omega  \left(
\begin{array}{cc}
0 & \mathrm{e}^{\ii \omega t} \\
\mathrm{e}^{-\ii \omega t} & 0
\end{array}
\right)\,,
\end{eqnarray}
with $\Omega$ the coupling strength and $\omega$ the frequency of the Hamiltonian oscillation (the detuning in the case of an atom, the frequency of the rotating magnetic field in the case of spin). The instantaneous eigenstates of this Hamiltonian are $\Ket{\pm}=(\Ket{1}\pm  \mathrm{e}^{-\ii \omega t}  \Ket{0})/\sqrt{2}$, corresponding to the eigenvalues $\pm \Omega$.

In our case, the adiabaticity condition previously recalled simply becomes: $|\Bra{+}\dot{H}_\mathrm{rot}\Ket{-}|/(2\Omega)^2 = \omega/(4\Omega) \ll 1$. Since the Hamiltonian is periodic, at time $t=T\equiv 1/\omega$ we have $H(T)=H(0)$, and if the system initial state $\Ket{\psi_0}$ is an eigenstate of $H(0)$, $\Ket{+(0)}$ or $\Ket{-(0)}$, at $t=T$ the system will return in its initial state, under an adiabatic evolution. Therefore a meaningful quantity to monitor is the survival probability of the initial state $\Ket{\psi_0}$: 
\begin{eqnarray}\label{eq:Fidelity_Def}
{\cal F}(t) = \mathrm{tr}(\rho(t)\KetBra{\psi_0}{\psi_0})\,.
\end{eqnarray}

\begin{figure}
\subfigure[]{\includegraphics[width=0.45\textwidth, angle=0]{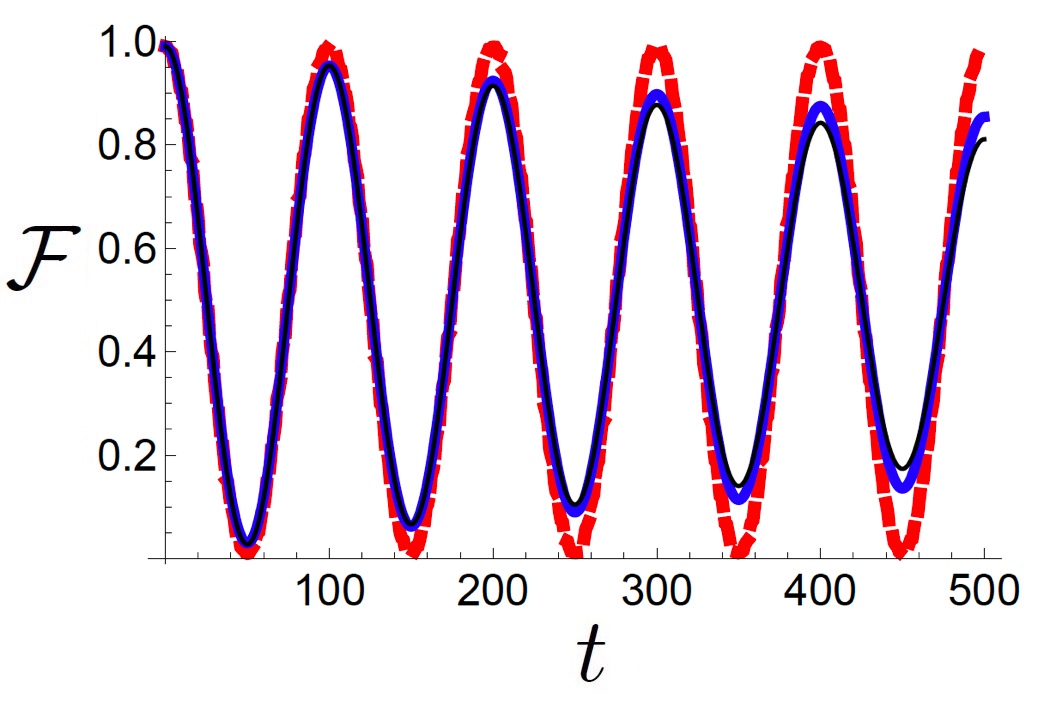}}
\subfigure[]{\includegraphics[width=0.45\textwidth, angle=0]{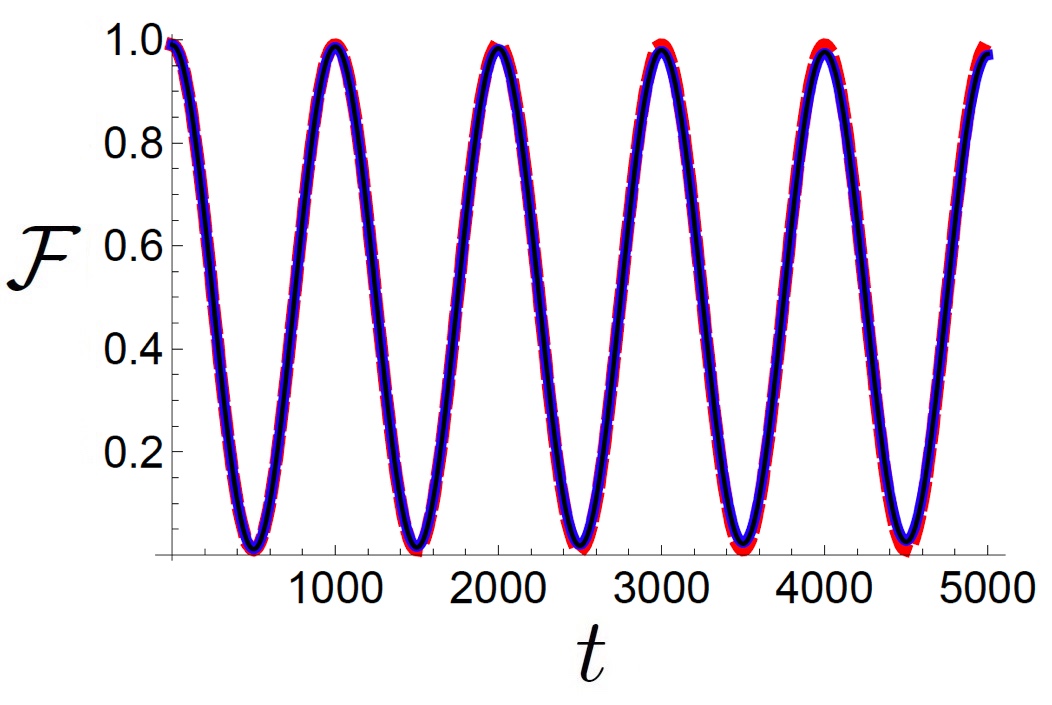}}
\caption{Population of the initial state ${\cal F}(t) = \mathrm{tr}(\rho(t)\KetBra{\psi_0}{\psi_0})$ as a function of time (in units of $1/\Omega$), under unitary evolution (red dashed bold line), and SEA evolutions for $\gamma/\Omega=0.5$ (blue solid line) and $\gamma/\Omega=2$ (black solid thin line). Here we have $\omega/\Omega=2\pi/100$ (a) and $\omega/\Omega=2\pi/1000$ (b). In both cases the initial state is $\rho(0)=(1-\lambda)\Ket{\psi_0}\Bra{\psi_0}+(\lambda/2) I$ with $\lambda=10^{-2}$ and $\Ket{\psi_0}=\Ket{+(0)}=(\Ket{0}+\Ket{1})/\sqrt{2}$.}\label{fig:2SS_Ad}
\end{figure}

\begin{figure}
\subfigure[]{\includegraphics[width=0.45\textwidth, angle=0]{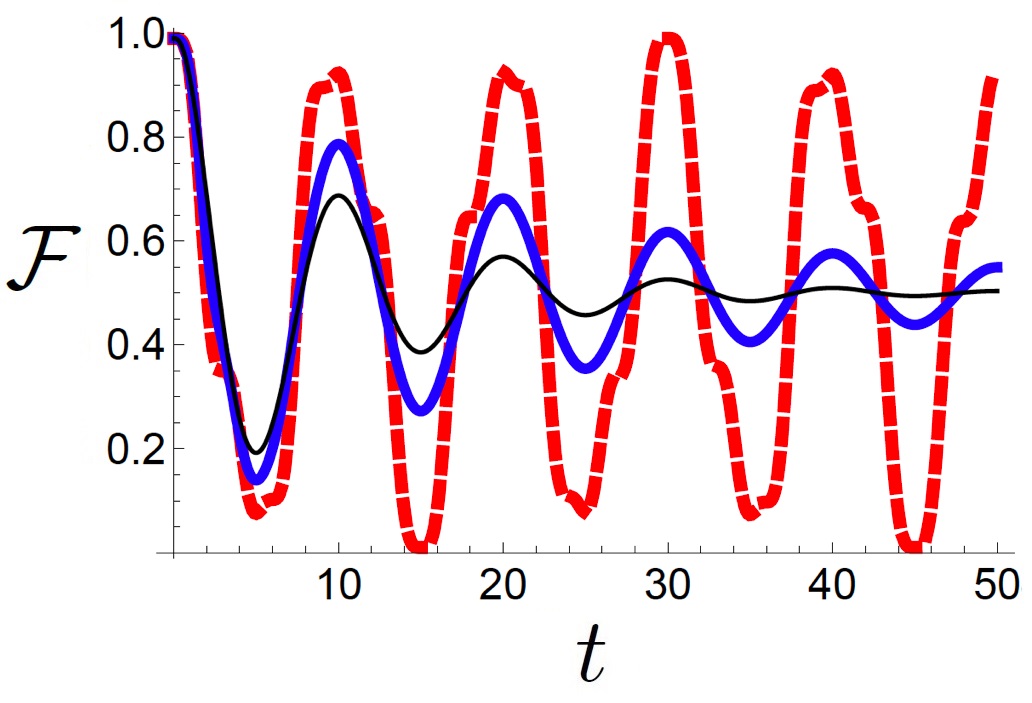}} \\ %
\subfigure[]{\includegraphics[width=0.45\textwidth, angle=0]{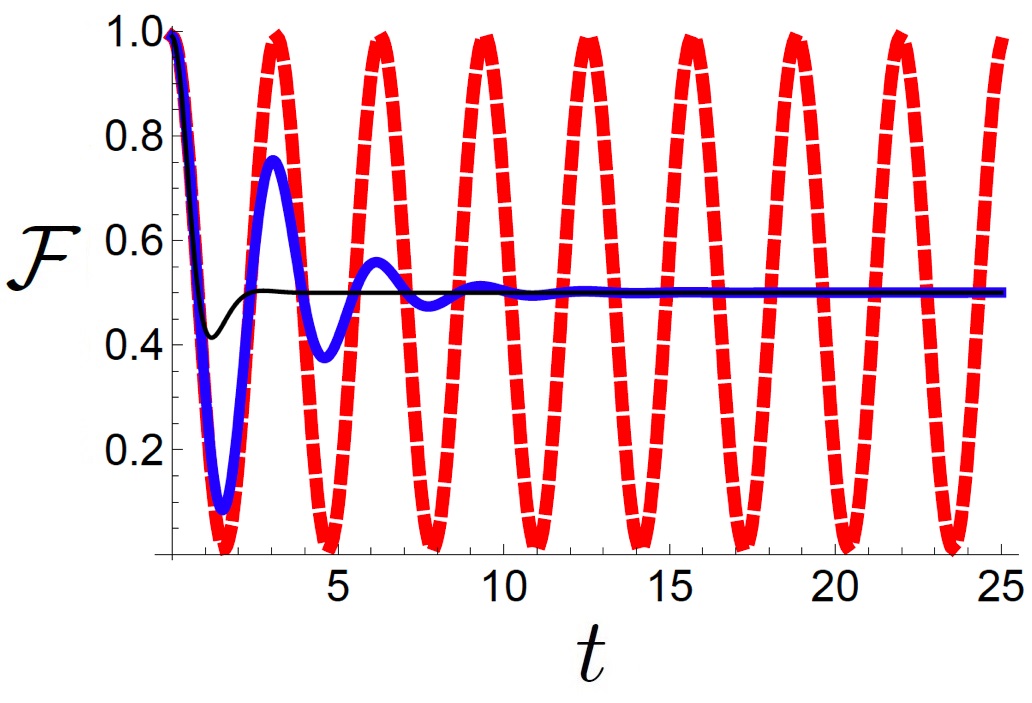}} %
\caption{Population of the initial state ${\cal F}(t) = \mathrm{tr}(\rho(t)\KetBra{\psi_0}{\psi_0})$ as a function of time (in units of $1/\Omega$), under unitary evolution (red dashed bold line), and SEA evolutions for $\gamma/\Omega=0.5$ (blue solid line) and $\gamma/\Omega=2$ (black solid thin line). In (a) we have $\rho(0)=(1-\lambda)\Ket{\psi_0}\Bra{\psi_0}+(\lambda/2) I$, $\Ket{\psi_0}=\Ket{+(0)}=(\Ket{0}+\Ket{1})/\sqrt{2}$ and $\omega/\Omega=2\pi/10$, while in (b) $\Ket{\psi_0}=\Ket{1}$ and $\omega/\Omega=2\pi/100$. In both cases $\lambda=10^{-2}$.} \label{fig:2SS_AdBreakdown}
\end{figure}

In Fig.~\ref{fig:2SS_Ad} it is shown an example of adiabatic evolution  when the system starts in a state $\rho(0)\approx\Ket{+(0)}\Bra{+(0)}$, an eigenstate of $H(0)$, in the absence and in the presence of entropy production and for different values of $\omega/\Omega$. As expected, it is well visible that discrepancies between unitary dynamics and SEA evolutions are very small, even after a long time. In particular, in Fig.~\ref{fig:2SS_Ad}b the ratio $\omega/\Omega$ is smaller than in Fig.~\ref{fig:2SS_Ad}a, which makes the adiabatic approximation more appropriate, which in turn diminishes the discrepancy between the unitary and the SEA evolutions (the three curves in Fig.~\ref{fig:2SS_Ad}b almost perfectly coincide).
In Fig.~\ref{fig:2SS_AdBreakdown} are shown two evolutions where the discrepancy between SEA evolutions and the corresponding unitary ones are very significant. Such discrepancies are easily understood in terms of violations of the relevant hypotheses. In Fig.~\ref{fig:2SS_AdBreakdown}a a non adiabatic evolution is considered. Indeed, $\omega/\Omega =2\pi/10$ does not guarantee adiabaticity. On the contrary, in Fig.~\ref{fig:2SS_AdBreakdown}b, the adiabatic condition can be assumed as fulfilled with $\omega/\Omega\sim 2\pi/100$ (it is the same ratio associated to the evolutions reported in Fig.~\ref{fig:2SS_Ad}a), but the system is initially prepared in a state $\rho(0)\approx\Ket{1}\Bra{1}$, which is far from any eigenstate of $H(0)$.

\begin{figure}
\includegraphics[width=0.45\textwidth, angle=0]{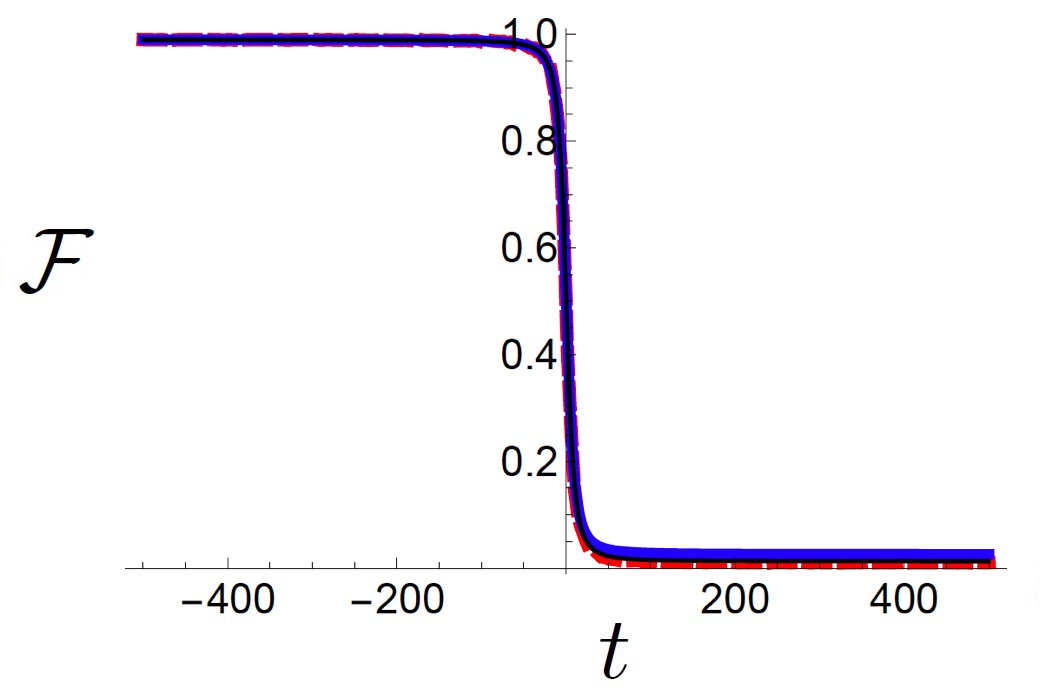}
\caption{Population of the initial state ${\cal F}(t) = \mathrm{tr}(\rho(t)\KetBra{\psi_0}{\psi_0})$ as a function of time (in units of $1/\xi$), under unitary evolution (red dashed bold line), and SEA evolutions for $\gamma/\xi=1$ (blue solid line) and $\gamma/\xi=10$ (black solid thin line). Relevant quantities are: $\rho(0)=(1-\lambda)\Ket{\psi_0}\Bra{\psi_0}+(\lambda/2) I$ with $\lambda=10^{-2}$, $\Ket{\psi_0}=\Ket{0}$, $\kappa/\xi^2=0.1$ and $\xi T=500$.} \label{fig:LZ}
\end{figure}

{\it Avoided crossing} --- As another example of adiabatic evolutions we consider the very archetypical and famous avoided-crossing scheme, corresponding to a TSS described by the following time-dependent Hamiltonian,
\begin{eqnarray}
H_\mathrm{LZ}(t) = \left(
\begin{array}{cc}
\kappa t & \xi \\
\xi & -\kappa t
\end{array}
\right)\,,
\end{eqnarray}
whose eigenvalues and eigenstates are $\pm\sqrt{\xi^2 + (\kappa t)^2}$, $\Ket{+}=\cos\theta \Ket{1}+\sin\theta\Ket{0}$ and $\Ket{-}=-\sin\theta \Ket{1}+\cos\theta\Ket{0}$, with $\theta=-\arctan((\kappa t - \sqrt{\xi^2 + (\kappa t)^2})/\xi)$. For $\kappa t \rightarrow \infty$ we have $\Ket{+} \rightarrow \Ket{1}$, while for $\kappa t \rightarrow -\infty$ we have $\Ket{+} \rightarrow \Ket{0}$ (assume $\kappa, \xi > 0$). This means that if the time evolution starts at a time $t=-T$ large enough to have $\kappa T \gg \xi$ and stop at $t=T$, and assuming that $\kappa$ is small enough to make the adiabatic approximation valid ($\kappa/\xi^2 \ll 1$), we have that the state $\Ket{1}$ is adiabatically mapped into $\Ket{0}$ and vice versa. If the evolution is not perfectly adiabatic, deviations can be evaluated through the remarkable result obtained independently by Landau, Zener, Majorana and Stueckelberg in the same year \cite{ref:LZ1}.  Since we are interested in demonstrating robustness of adiabatic evolutions against the SEA non-unitary contributions to the dynamics, we will consider cases where the adiabatic condition is satisfied. In Fig.~\ref{fig:LZ} it is shown the population of the initial state $\Ket{0}$ in the state $\rho(t)$: this quantity starts being (very close to) $1$, since the initial state is $\rho(0)\approx\Ket{0}\Bra{0}$, and changes to the point of vanishing, meaning that an almost complete transition $\Ket{0}\Bra{0} \rightarrow \Ket{1}\Bra{1}$ occurs. The three curves corresponding to the three different values of $\gamma/\xi=0,1,10$ almost coincide, demonstrating again that the adiabatic following of the relevant eigenstate of the Hamiltonian is robust against the SEA pushing. Starting with the state $\Ket{1}$ a very similar plot (not reported here) is obtained.

\section{Discussion}\label{sec:conclusion}

Summarizing, we have considered the Steepest Entropy Ascent approach applied to TSSs, bringing to light the very peculiar property that when the Hamiltonian is time independent the populations of the eigenstates of the Hamiltonian do not change at all during the evolution, in spite of the non unitary contribution to the time evolution. Coherences of course do change. Subsequently, we have analyzed the problem of how to obtain SEA master equations in the presence of time dependent Hamiltonians. Though the approach itself has been used in previous works with time-dependent Hamiltonians, we have presented some arguments supporting exploitation of SEA approach in this case with a master equation of the form given in \eqref{eq:SEAME_Canonical}. In fact, the recipe to write down SEA master equations in the presence of time dependent Hamiltonians can be better understood by considering that for every $t$ a time window $(t, t+\tau)$ exists short enough to guarantee that the system does not distinguish between a time dependent Hamiltonian and a time independent one. Indeed, if the relevant Hilbert space $\cal H$ is finite-dimensional, the time dependent Hamiltonian $H(t)$ has a finite set of eigenvalues and eigenstates, say $\epsilon_0(t),... \epsilon_{N-1}(t)$ and $\Ket{\phi_0(t)},... \Ket{\phi_{N-1}(t)}$ (with $N=\mathrm{dim}{\cal H}$), and for every $\eta>0$ it is possible to find a $\tau$ such that $\forall k=0,1,..., N-1$ and $\forall t' \in (t,t+\tau)$\, $|\epsilon_k(t')-\epsilon_k(t)| < \eta$ and $1-|\BraKet{\phi_k(t)}{\phi_k(t')}|<\eta$, which allows for assuming that $H(t)$ can be considered as if it was constant. On the contrary, if the the dimension of the Hilbert space is infinite, not necessarily such a value $\tau$ can be found that allows for having the discrepancies of an infinite set of function to be smaller than $\eta$. Anyway, in our case the argument is surely valid.

Once the SEA master equation has been derived, we have analyzed it in the adiabatic limit, i.e., for Hamiltonians which change very slowly. We have provided reasons for expecting a certain robustness of adiabatic SEA evolutions of TSSs. In the examples we have considered (a TSS in a rotating field and an avoided crossing of two levels) the predicted robustness against SEA non unitary part of dynamics is well visible.

\appendix

\section{SEA Master Equation for Canonical Contact Conditions}\label{app:SEAME_Canonical}

In this appendix we recall how to derive the SEA master equation according to the approach reported in Refs.~\cite{ref:Beretta2007, ref:Beretta2009, ref:Beretta2006, ref:Beretta2014}.

First, one introduce the scalar product between two operators defined by $(A | B) \equiv \mathrm{tr}(A B^\dag + A^\dag B)/2$. Second, one finds that the variation of the entropy functional  $s(\rho)\equiv \mathrm{tr}(\rho\log\rho)$ is $\mathrm{d} s(\rho)/\mathrm{d}t=\mathrm{tr}(({I}+\log\rho)\dot{\rho})=\mathrm{tr}(({I}+\log\rho)(\rho E+E^\dag\rho))=2 ( \sqrt{\rho} E | \sqrt{\rho}({I}+\log\rho))$. (Here ${I}$ is the identity of the relevant Hilbert space). It is then obvious that by choosing $E=({I}+\log\rho)$ one would move along the path of maximum of entropy production, but since we need to satisfy the relevant constraints, we have to remove from the \lq vector\rq\, ${I}+\log\rho$ the components which can produce changes of the constants of motion. Now observe that, given an Hermitian operator $X$, the relevant expectation value changes with the following rate: $\mathrm{d}\langle X \rangle_\rho /\mathrm{d}t = \mathrm{d}\,\mathrm{tr}(\rho X)/\mathrm{d}t = 2 (\sqrt{\rho} X | \sqrt{\rho} E )$. Therefore, in order to obtain the conservation of an operator $X$ it is necessary and sufficient to impose that the scalar product between $\sqrt{\rho}E$ and $\sqrt{\rho}X$ is zero. With the canonical contact conditions we need to prevent probability loss and require energy conservation. Such two conditions correspond to impose the conservation of the two operators ${I}$ and $H$, so that $\sqrt{\rho} E=\sqrt{\rho} \,{I} + \sqrt{\rho} \log\rho +\alpha' \sqrt{\rho}  \,{I} + \beta \sqrt{\rho} H = \sqrt{\rho}  \log\rho +\alpha \sqrt{\rho} \,{I} + \beta \sqrt{\rho} H$, where $\alpha'$, $\alpha=\alpha'+1$ and $\beta$ are suitable coefficients. Such coefficients can be straightforwardly found through a Gram determinant. After introducing the symbol $( X | Y )_\rho \equiv ( \sqrt{\rho} X | \sqrt{\rho} Y )$, we can write:
\begin{eqnarray}
\nonumber 
\sqrt{\rho} E &=&
\mathrm{det} 
\left(
\begin{array}{ccc}
-\sqrt{\rho}\log\rho & \sqrt{\rho} \,{I} & \sqrt{\rho} H \\
& & \\
-( \log\rho | {I} )_\rho & ( {I} | {I} )_\rho  & ( {I} | H )_\rho  \\ 
& & \\
-( \log\rho | H )_\rho & ( H | {I} )_\rho  & ( H | H )_\rho  
\end{array}
\right) / \\ 
\nonumber \\
&& \mathrm{det} \left(
\begin{array}{cc}
( {I} | {I} )_\rho  & ( {I} | H )_\rho  \\ 
 & \\
( H | {I} )_\rho  & ( H | H )_\rho  
\end{array}
\right)\,,
\end{eqnarray} 
or, equivalently, 
\begin{eqnarray}
\label{app:determinant}
\nonumber 
E &=&
\mathrm{det} 
\left(
\begin{array}{ccc}
-\log\rho & {I} & H \\
& & \\
s(\rho) & 1  & \langle H\rangle_\rho  \\ 
& & \\
-\langle \log\rho H \rangle_\rho & \langle H \rangle_\rho  & \langle H^2 \rangle_\rho  
\end{array}
\right) / \\ 
\nonumber \\
&& \mathrm{det} \left(
\begin{array}{cc}
1  & \langle H \rangle_\rho    \\ 
 & \\
\langle H \rangle_\rho    & \langle H^2 \rangle_\rho    
\end{array}
\right)\,,
\end{eqnarray} 
which, after some algebra, leads to \eqref{eq:SEAME_Canonical}.

\section{Stationarity of The Canonical State}\label{app:CanonicalState}

In this appendix we prove that every canonical state is a stationary state for a master equation of the form in  \eqref{eq:SEAME_Canonical}. A simple way to prove this assertion is to observe that when $\rho$ is a canonical state $\rho = \omega(\beta)$   both $[H, \omega(\beta)]$ and the determinant in \eqref{app:determinant} are zero. The former assertion is trivial, since $\omega(\beta)$ is a function of $H$. The latter can be straightforwardly proven by observing what the first column of the (numerator) determinant becomes. After introducing $C(\beta)=\mathrm{tr} e^{-\beta H}$, we find $-\log\omega(\beta) = \beta H + \log C(\beta) \, {I}$, $s(\omega(\beta)) =  \beta\langle H\rangle_\omega + \log C(\beta)$, $ -\langle \log\omega H\rangle_\omega= \beta\langle H^2\rangle_\omega+ \log C(\beta)$. Therefore, the first column turns out to be a linear combination of the second (with coefficient $\log C(\beta)$) and the third (with coefficient $\beta$), which implies that $E(\rho)$ vanishes.

Restrictions of a canonical state to every subspace generated by eigenstates of the Hamiltonian are stationary states as well. When a restriction is considered, possible singularities can appear in the numerator and in the $2\times 2$ determinant in the denominator. Nevertheless, all possible singularities are adequately compensated when the complete terms appearing in the master equation, $\rho E$ and $E^\dag\rho$, are considered, leading to $\omega_{\hat{P}}(\beta) E = 0$ and $E^\dag \omega_{\hat{P}}(\beta) = 0$.

\section{Adiabatic Approximation}\label{app:Adiabaticity}

In this appendix we briefly recall the essential aspects of the adiabatic approximation. We will follow the line of Ref.\cite{ref:Wilczeck} instead of that of Ref.\cite{ref:Messiah}. For the sake of simplicity we will restrict our analysis to the case of non degenerate spectrum of the Hamiltonian.

Consider a system governed by a time-dependent Hamiltonian $H(t)$ with eigenvalues $\epsilon_k(t)$ and eigenstates $\Ket{\phi_k(t)}$. Let $\Ket{\psi(t)}$ denotes the state of the system, which can be expanded in terms of the instantaneous eigenstates of the Hamiltonian: 
\begin{eqnarray}
\Ket{\psi(t)} = \sum_k a_k(t) \,\ee^{-\ii \int_o^t \epsilon_k(s)\mathrm{d}s} \, \Ket{\phi_k(t)}\,.
\end{eqnarray} 
The insertion of such expansion in the Schr\"odinger equation gives rise to the following equation for the coefficients:
\begin{eqnarray}
\nonumber
\dot{a}_n(t) = -\sum_k  a_k(t) \,\ee^{-\ii \int_o^t (\epsilon_k(s)-\epsilon_n(s)) \mathrm{d}s} \, \langle{\phi_n(t)}|{\dot{\phi}_k(t)}\rangle\,, \\
\end{eqnarray} 
which can be rearranged in the following way:
\begin{eqnarray}\label{eq_eqforan}
\nonumber
\dot{a}_n(t) = &-& a_n(t) \, \langle{\phi_n(t)}|{\dot{\phi}_n(t)}\rangle\\
\nonumber
 &+&\sum_{k\not=n}  a_k(t) \, \ee^{-\ii \int_o^t (\epsilon_k(s)-\epsilon_n(s)) \mathrm{d}s} \, \\
 && \times \frac{\langle \phi_n(t) | \dot{H}(t) | \phi_k(t) \rangle}{\epsilon_n(t)-\epsilon_k(t)} \,,
\end{eqnarray} 
where we have used the following properties: %
(i)  $\langle{\dot{\phi}_n(t)}|{\phi_k(t)}\rangle + \langle{\phi_n(t)}|{\dot{\phi}_k(t)}\rangle = 0$, which comes from $0=\partial_t \langle{\phi_n(t)}|{\phi_k(t)}\rangle$; %
(ii) $\langle{\phi_n(t)}|{\dot{\phi}_k(t)}\rangle= - \langle \phi_n(t) | \dot{H}(t) | \phi_k(t) \rangle / (\epsilon_n(t)-\epsilon_k(t))$ for $k\not=n$, which comes from $0=\partial_t \Bra{\phi_j(t)} H(t) \Ket{\phi_l(t)} =$ $\langle \dot\phi_j(t) | H(t) | \phi_l(t) \rangle +   \langle \phi_j(t) | \dot{H}(t) | \phi_l(t) \rangle + \langle \phi_j(t) | H(t) | \dot\phi_l(t) \rangle$  $= \langle \phi_j(t) | \dot{H}(t) | \phi_l(t) \rangle  +   (\epsilon_j(t)-\epsilon_l(t)) \langle \phi_j(t) | \dot\phi_l(t) \rangle$, which for $\epsilon_j(t)-\epsilon_l(t)\not=0$ --- coinciding with the condition $j\not=l$ in the case of non degenerate spectrum --- can be solved for $\langle \phi_j(t) | \dot\phi_l(t) \rangle$. 

When the condition  $|\langle \phi_n(t) | \dot{H}(t) | \phi_k(t) \rangle / (\epsilon_n(t)-\epsilon_k(t))| \ll |\epsilon_n(t)-\epsilon_k(t)|$ is satisfied (i.e., the instantaneous coupling strengths are much smaller than the instantaneous frequencies in the phase factors) all the terms of the right-hand side of \eqref{eq_eqforan} with $k\not=n$ can be neglected and each $a_n$ is essentially given by an exponential. Moreover, since from property (i) specialized to the case $k=n$ comes that $\langle{\phi_n(t)}|{\dot{\phi}_n(t)}\rangle$ is a pure imaginary, it turns out that the modulus of the coefficient $a_n(t)$ does not change.

The generalization of this calculation to the case of an Hamiltonian with degenerate subspaces implies that transitions between states belonging to different instantaneous eigenspaces of the Hamiltonian are forbidden, provided the matrix elements between such states (belonging to different eigenspaces) are much smaller than the squares of the relevant Bohr frequencies. Of course, transitions between states belonging to the same eigenspace are allowed anyway. Indeed, property (ii) cannot be obtained if two states belonging to the same subspace are considered, since in that case $\epsilon_n-\epsilon_k=0$.

\section*{Acknowledgements}

The Author thanks Anna Napoli for carefully reading the manuscript.


\begin{thebibliography}{99}
\bibitem{ref:footnote} So far, the SEA principle hasn't been neither validated nor invalidated. This work is to be meant as an exploration of the consequences of the SEA approach in the presence of slowly varying Hamiltonians.
\end{thebibliography}
\end{document}